\documentclass[twocolumn,aps,superscriptaddress,showpacs,nofootinbib]{revtex4}

\usepackage{amssymb}
\usepackage{amsmath}
\usepackage{graphicx}
\usepackage[normalem]{ulem}
\usepackage[dvips]{color}

\usepackage[normalem]{ulem}  
\usepackage[dvips]{color} 

\renewcommand\sout{\bgroup \color{red} \ULdepth=-.5ex \ULset}

\begin{document}
\title{Chemical freeze-out in relativistic heavy-ion collisions}

\author{Jun Xu}
\email{xujun@sinap.ac.cn}
\affiliation{Shanghai Institute of Applied
Physics, Chinese Academy of Sciences, Shanghai 201800, China}

\author{Che Ming Ko}
\email{ko@comp.tamu.edu}
\affiliation{Cyclotron Institute and
Department of Physics and Astronomy, Texas A$\&$M University,
College Station, Texas 77843, USA}

\date{\today}

\begin{abstract}
One surprising result in relativistic heavy-ion collisions is that
the abundance of various particles measured in experiments is
consistent with the picture that they reach chemical equilibrium at
a temperature much higher than the temperature they freeze out
kinetically. Using a multiphase transport model to study particle
production in these collisions, we find that the above result is due
to the constancy of the entropy per particle during the evolution of
the hadronic matter from the chemical to the kinetic freeze-out.  We
further use a hadron resonance gas model to illustrate the result
from the transport model study.
\end{abstract}


\maketitle

The statistical model, which assumes that the abundance of particles
produced in relativistic heavy-ion collisions or their chemical
freeze-out is determined before the hadronic matter freezes out
kinetically, has been very successful in describing the yields of
particles measured in experiments. For heavy-ion collisions at the
top energy of the Relativistic Heavy Ion Collision (RHIC) and the
Large Hadron Collider (LHC)~\cite{STAR,PHENIX,LHC}, the chemical
freeze-out temperature extracted from the experimental data using
the statistical model is around $160\sim 170$ MeV and is similar to
the critical temperature of the hadron-quark phase transition from
lattice QCD calculation~\cite{STAR,PHENIX,LHC,LQCD}. Although it is
not known if the chemical freeze-out temperature in the collision
energy regime of the RHIC Beam-Energy Scan (BES) program, which is
around $140 \sim 160$ MeV, coincides with the hadron-quark phase
boundary, relativistic heavy-ion collisions provide the only
possibility to map out experimentally the phase diagram of the
strong-interacting matter in the temperature ($T$) - chemical
potential ($\mu$) plane~\cite{lrp}. Since the chemical freeze-out
temperature is much higher than that for the kinetic freeze-out,
which is around $100 \sim 120$ MeV from the blast-wave fit to
experimental transverse momentum/mass spectra~\cite{Kum14,STAR17}
and slightly decreases with increasing collision energy, to maintain
the same relative abundance among various particles requires
non-unity values for their fugacities as the hadronic matter expands
and cools from the chemical freeze-out temperature $T_{ch}$ to the
kinetic freeze-out temperature $T_{kin}$.

To study the origin of early chemical freeze-out, we use a
multiphase transport model (AMPT)~\cite{AMPT}, which has been widely
used as a theoretical tool or an event generator for relativistic
heavy-ion collisions. In this model, the initial-state information
is generated by the Heavy-Ion Jet INteraction Generator (HIJING)
model~\cite{hijing}, which is an extension of the PYTHIA
model~\cite{pythia} for proton-proton collisions to nucleus-nucleus
collisions. In the string melting version of AMPT, which is employed
in the present study, all the hadrons produced from HIJING are
converted into their valence quarks, and the dynamics of these
quarks is described by Zhang's Parton Cascade (ZPC)
model~\cite{zpc}. The parton scattering cross section is set to be
1.5 mb, which has been shown to reproduce reasonably well the
experimentally measured collective flow in relativistic heavy-ion
collisions~\cite{Xu11}. The freeze-out of partons is determined by
their last scatterings. Hadrons are then formed from a spatial
coalescence mechanism with their species determined by the flavors
and invariant mass of their constituent quarks. To reproduce
reasonably the empirical energy density and/or temperature near
hadronization, we remove the additional hadron formation time of 0.7
fm/c, which was previously introduced to reproduce the hadron
multiplicities at midrapidities~\cite{AMPT}, except the usual one
given by $E/m_T^2$ in terms of the energy $E$ and transverse mass
$m_T$ of the hadron. The evolution of the hadronic phase is
described by a relativistic transport (ART) model~\cite{art} that
includes various hadron species as well as elastic, inelastic, and
decay channels among these hadrons. The empirical energy dependence
of the scattering cross sections, the Breit-Wigner mass distribution
of resonances, and the mass dependence of the decay width are
properly taken into account by satisfying the detailed balance
condition, as described in Refs.~\cite{art,AMPT}. The frequent
scatterings among hadrons are helpful to maintain the thermal
equilibrium in the hadronic phase. In the present work we study
central Au+Au collisions at center-of-mass energy
$\sqrt{s_{NN}}=200$ and 7.7 GeV, corresponding to the top RHIC
energy and a typical lower collision energy at the RHIC-BES program
where the partonic phase is less dominant. A total number of 100 and
1000 events are generated at $\sqrt{s_{NN}}=200$ GeV and $7.7$ GeV,
respectively.

\begin{figure}[h]
\centerline{\includegraphics[scale=0.8]{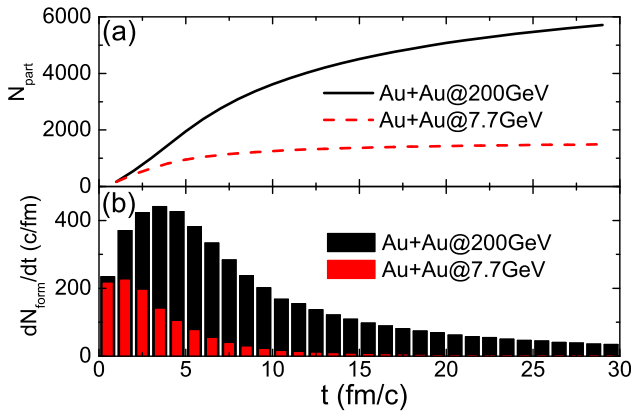}} \caption{(Color
online). Time evolution of the total number of particles (a) and
their formation rate (b) in the hadronic phase of central Au+Au
collisions at $\sqrt{s_{NN}}=200$ and 7.7 GeV. } \label{number}
\end{figure}

To give a general picture for particle production in the hadronic
phase, we show in Fig.~\ref{number} the time evolution of the total
number of particles and their formation rate in central Au+Au
collisions from the AMPT model. It is seen that the total number of
hadrons increases with time due to the continuous production of
hadrons from quark coalescence as well as inelastic and decay
channels in the hadronic phase. The different production times of
hadrons from quark coalescence are a result of different quark
freeze-out times from their last scatterings. Although the final
hadron number is much larger at 200 GeV than at 7.7 GeV, hadrons are
mostly formed at earlier times at lower collision energies due to
the shorter lifetime of the partonic phase.

\begin{figure}[h]
\centerline{\includegraphics[scale=0.8]{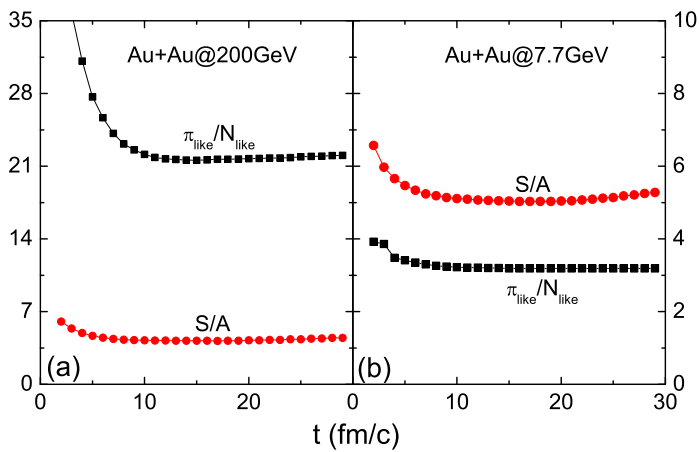}} \caption{(Color
online). Time evolution of the entropy per particle ($S/A$) and the
effective pion/nucleon ratio ($\pi_{\rm like}/N_{\rm like}$) in the
hadronic phase of central Au+Au collisions at $\sqrt{s_{NN}}=200$
(a) and 7.7 GeV (b). } \label{ratio}
\end{figure}

The continuous production of hadrons makes it not possible to define
unambiguously the chemical freeze-out time. Instead, we consider the
phase-space distribution in the hadronic matter at different times.
The resulting time evolution of the effective pion/nucleon ratio is
shown by filled squares in Fig.~\ref{ratio} for collisions at both
200 GeV (left panel) and 7.7 GeV (right panel). In calculating the
effective pion/nucleon ratio, we include those from the strong
decays of resonances. Specifically, the pion-like particles are
pions, $\Delta$ and $N^*$ resonances, $\rho$ mesons, $\omega$
mesons, $\eta$ mesons, $K^*$ mesons, and their antiparticles, while
the nucleon-like particles are nucleons and their resonances
$\Delta$ and $N^*$.

Since entropy is a thermodynamical quantity that contains
information on both the number of degrees of freedom in a system and
its degree of thermal equilibration, it is of interest to study this
quantity in the hadronic phase of AMPT. Like the viscous effect in
hydrodynamic description of heavy-ion
collisions~\cite{Hos85,Dan85,Hei96,Mur02}, entropy is produced in
AMPT from scattering and production of particles. However, more
entropy is expected to be produced in hadronic matter than in
partonic matter because of its significantly larger specific
viscosity~\cite{De09}.  Based on the thermal model, it has been
shown that the entropy per particle in intermediate-energy heavy-ion
collisions is related to the deuteron/proton
ratio~\cite{Sie79,Aic87}. Therefore, the entropy per particle and
relative abundance of particle species in heavy-ion collisions are
believed to be related, with the latter often used to extract the
information about the former~\cite{Dos88,Kuh93}. However, the
relation between the entropy per particle and the relative abundance
of particles has so far been studied under the assumption that the
system is in chemical equilibrium.

\begin{figure}[h]
\centerline{\includegraphics[scale=0.8]{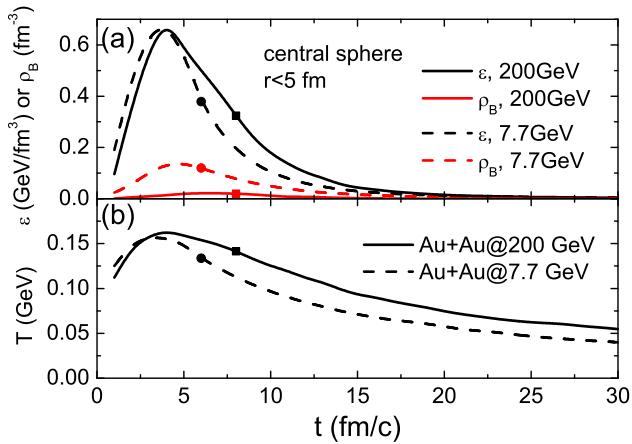}} \caption{(Color
online). Time evolution of the energy density $\epsilon$ and the net
baryon density $\rho_B$ (a) in the central sphere of radius $r=5$ fm
in the hadronic phase of central Au+Au collisions at
$\sqrt{s_{NN}}=200$ and 7.7 GeV as well as the extracted temperature
(b) from the hadron resonance gas model. Filled squares and circles
indicate the time after which the effective pion/nucleon ratio and
the entropy per particle remain approximately constant.}
\label{tebt}
\end{figure}

In the AMPT, the entropy can be calculated using the phase-space
distributions $f_i({\bf x},{\bf p})$ of particle species $i$ as in Refs.~\cite{Ber81,Gud85}, i.e.,
\begin{eqnarray}
S= -\sum_i g_i \int\frac{d^3{\bf x}d^3{\bf p}}{(2\pi)^3}[f_i\ln f_i \pm (1 \mp f_i)\ln(1 \mp f_i)],\label{s}
\end{eqnarray}
with $g_i$ being the spin degeneracy of a hadron. For $f_i$ in the
above equation, it is evaluated from counting the number of
particles in a local six-dimensional phase-space cell. The size of
the phase-space cell is carefully chosen to be small but include
sufficient number of particles after averaging over all
events~\cite{Ber81}. We use spherical coordinates for both position
and momentum and divide the phase space into cells. Because of the
symmetry in the azimuthal angle for central collisions, the
dimension in the phase-space coordinate is reduced. We further
introduce a momentum cut of $|p|<4$ GeV/c in obtaining the results
in Fig.~\ref{ratio} since low-momentum particles dominate at
midrapidities. It is seen from Fig.~\ref{ratio} that the entropy per
particle and the effective pion/nucleon ratio decrease at the early
stage of the hadronic phase, and both become approximately constants
at later stage. This behavior is similar to that observed in
Ref.~\cite{Gud85}, where both the pion/nucleon ratio and the entropy
per particle remain constant after the most compressed stage in
intermediate-energy heavy-ion collisions. Although this result seems
to be consistent with the argument that the entropy per particle is
related to the relative abundance of particle species at and after
chemical freeze-out~\cite{Sie79,Aic87,Dos88,Kuh93}, we will show
later that the constant entropy per particle and the constant
effective pion/nucleon ratio cannot be achieved simultaneously
unless the system becomes out of chemical equilibrium in later
stage. However, we can still define the chemical freeze-out time,
which is about 8 fm/c at $\sqrt{s_{NN}}=200$ GeV and about 6 fm/c at
7.7 GeV, indicated in Fig.~\ref{tebt} by filled squares and circles,
respectively, after which the effective pion/nucleon ratio remains
essentially unchanged. It is also seen that it takes a longer time
to reach chemical freeze-out at 200 GeV than at 7.7 GeV, and this is
due to the earlier production of hadrons at lower collision energies
as shown in Fig.~\ref{number}.

Results from AMPT, which does not assume thermal and chemical
equilibrium, can be qualitatively understood in terms of the hadron
resonance gas model that includes all the hadron species in the
ART/AMPT model. For such a system containing non-interacting
hadrons, the number density $\rho$, the energy density $\epsilon$,
and the entropy density $s$ can be respectively expressed
as~\cite{sta,And06}
\begin{eqnarray}
\rho &=& \sum_i g_i \int \frac{d^3p}{(2\pi)^3}f_i ,\label{rho}\\
\epsilon &=& \sum_i g_i \int  \frac{d^3p}{(2\pi)^3}f_i\sqrt{m_i^2+p^2},\\
s &=& -\sum_i g_i \int\frac{d^3p}{(2\pi)^3}[f_i\ln f_i \pm (1 \mp f_i)\ln(1 \mp f_i)],\label{s}
\end{eqnarray}
with $m_i$ being the mass of hadron species $i$. The occupation
number of hadrons in momentum is given by
\begin{equation}
f_i = \frac{1}{\lambda_i\exp[(\sqrt{m_i^2+p^2}-\mu_i)/T]\pm 1},
\end{equation}
where $\mu_i = B_i\mu_B + C_i\mu_c$ is the chemical potential of
particle species $i$, with $B_i$ and $C_i$ being its respective
baryon and charge numbers, $\mu_B$ and $\mu_c$ being its respective
baryon and charge chemical potentials, and $\lambda_i$ is the
fugacity that takes into account the possible violation of chemical
equilibrium. In the above equations, the upper signs are for
Fermions and the lower signs are for Bosons. The charge chemical
potential $\mu_c$ is determined by the charge conservation condition
$\rho_c/\rho_B=79/197$ for Au+Au collisions, with the charge density
$\rho_c$ and the net baryon density $\rho_B$ calculated similarly as
Eq.~(\ref{rho}). The entropy per particle $S/A$ from the hadron
resonance gas model is $s/\rho$. Including the same pion-like and
nucleon-like particles as in the AMPT calculations and assuming that
they are in separate chemical equilibrium, their fugacities can be
written as $\lambda_i=\lambda_{N}^{z_N} \lambda_{\pi}^{z_\pi}$,
where $z_N$ ($z_\pi$) is the effective nucleon (pion) number. For
instance, we have $z_N=1$ and $z_\pi=1$ for $\Delta$ resonances, and
$z_N=0$ and $z_\pi=2$ for $\rho$ mesons.

We first evaluate the energy density $\epsilon$ and the net baryon
density $\rho_B$ in the central sphere of the hadronic phase in
AMPT, and the results are shown in the upper panel of
Fig.~\ref{tebt}. We then use the hadron resonance gas model to
obtain from $\epsilon$ and $\rho_B$ the temperature, and its time
evolution is shown in the lower panel of Fig.~\ref{tebt}. The
highest energy density of about 0.65 GeV/fm$^3$ is seen to
correspond to the highest temperature of about 152 MeV, which are
similar for collisions at both $\sqrt{s_{NN}}=200$ and 7.7 GeV. The
chemical freeze-out temperature $T_{\rm ch}$ is 141 MeV at 200 GeV
at about $t=8$ fm and 134 MeV at 7.7 GeV at about $t=6$ fm, which
are slightly lower than those extracted from the experimental data
based on the statistical model~\cite{STAR,PHENIX,
Cel06,And10,STAR17}.

\begin{figure}[h]
\centerline{\includegraphics[scale=1.0]{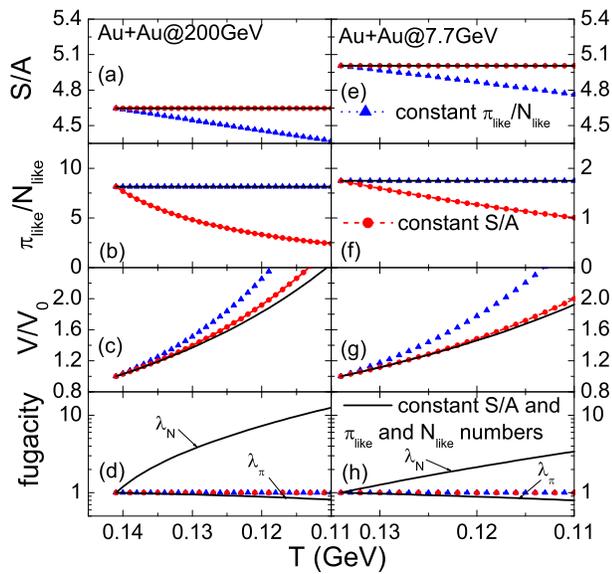}}
\caption{(Color online). Evolution of the entropy per particle
(first row), the effective pion/nucleon ratio (second row), the
volume expansion rate (third row), and the fugacities for pion-like
and nucleon-like particles (bottom row) as the hadron resonance gas
cools with its volume expansion rate and/or the values of particle
fugacities varied to keep a constant effective pion/nucleon ratio
(filled triangles), a constant entropy per particle (filled
circles), or the constancy of both (solid lines). See text for
details.  } \label{thermal}
\end{figure}

Using an initial state that is in chemical equilibrium with its
thermodynamical properties given by filled symbals in
Fig.~\ref{tebt}, we have studied the evolution of the entropy per
particle, the effective pion/nucleon ratio, the volume expansion
rate, and the fugacities of pion-like and nucleon-like particles as
the system cools down in a hadron resonance gas model from three
different scenarios. In the first scenario shown by filled triangles
in Fig.~\ref{thermal}, we find that although it is possible to keep
the constant effective pion/nucleon ratio by adjusting the volume
expansion rate of the hadronic matter, the entropy per particle
decreases as the system cools down. In the second scenario shown by
filled circles in Fig.~\ref{thermal}, we find that it is possible to
keep the entropy per particle a constant with a slower volume
expansion rate, but the effective pion/nucleon ratio decreases. To
keep both the entropy per particle and the effective pion/nucleon
ratio constant as in Fig.~\ref{ratio} from the AMPT model, it is
necessary to introduce non-unity values for the fugacity parameters.
Here we assume that all pion-like and nucleon-like particles are not
in chemical equilibrium while their effective numbers remain the
same after initial chemical freeze-out. It is seen from the solid
lines in Fig.~\ref{thermal} that in this third scenario the system
becomes increasingly out of chemical equilibrium during its
expansion and cooling as indicated by increasing nucleon and
decreasing pion fugacities away from unity.

To summarize, we have investigated the chemical freeze-out condition
in relativistic heavy-ion collisions based on a multiphase transport
model. Despite the continuous production of hadrons, the chemical
freeze-out time can be determined when the effective pion/nucleon
ratio becomes a constant. The latter is also found to be accompanied
by a constant entropy per particle. Starting from the chemical
freeze-out state in the AMPT model, we have further studied the
expansion and cooling of the system using the hadron resonance gas
model, and found that only the scenario of an expanding and cooling
hadronic matter with non-unity fugacities can lead to both constant
entropy per particle and effective pion/nucleon ratio. Our study
shows that after chemical freeze-out in relativistic heavy-ion
collisions, the system is no longer in chemical equilibrium, but the
statistical model can still be used to extract the temperature and
chemical potential at chemical freeze-out since the relative
abundances of particle species remain constant during later hadronic
evolution. The present study thus helps clarify our understanding of
chemical freeze-out in relativistic heavy-ion collisions, and
validate the use of the statistical model in mapping out the phase
diagram of the strong-interacting matter from relativistic heavy-ion
collisions.

We thank Chen Zhong for maintaining the high-quality performance of the computer facility.
The work of JX was supported by the Major State Basic Research
Development Program (973 Program) of China under Contract Nos.
2015CB856904 and 2014CB845401, the National Natural Science
Foundation of China under Grant Nos. 11475243 and 11421505, the
"100-talent plan" of Shanghai Institute of Applied Physics under
Grant Nos. Y290061011 and Y526011011 from the Chinese Academy of
Sciences, the Shanghai Key Laboratory of Particle Physics and
Cosmology under Grant No. 15DZ2272100, and the "Shanghai Pujiang
Program" under Grant No. 13PJ1410600, while that of CMK was supported by the US Department of Energy under Contract No. DE-SC0015266
and the Welch Foundation under Grant No. A-1358.


\begin{thebibliography}{99}

\bibitem{STAR} J. Adams {\it et al.} [STAR Collaboration], Nucl. Phys. A \textbf{757}, 102 (2005).
\bibitem{PHENIX} K. Adcox {\it et al.} [PHENIX Collaboration], Nucl. Phys. A \textbf{757}, 184 (2005).

\bibitem{LHC} B. Abelev {\it et al.} [ALICE Collaboration], Phys. Rev. C \textbf{88}, 044910 (2013).

\bibitem{LQCD} F. Karsch, Lect. Notes Phys. \textbf{583}, 209 (2002).

\bibitem{lrp} Y. Akiba {\it et al.}, arXiv: 1502.02730 [nucl-th].

\bibitem{STAR17} L. Adamczyk {\it et al.}, arXiv: 1701.07065 [nucl-ep].
\bibitem{Kum14} L. Kumer, Nucl. Phys. A \textbf{931}, 1114 (2014).

\bibitem{AMPT} Z. W. Lin, C. M. Ko, B. A. Li, B. Zhang, and S. Pal, Phys. Rev. C {\bf 72}, 064901 (2005).
\bibitem{hijing} X. N. Wang and M. Gyulassy, Phys. Rev. D {\bf 44}, 3501 (1991).
\bibitem{pythia} http://home.thep.lu.se/$\sim$torbjorn/Pythia.html
\bibitem{zpc} B. Zhang, Comput. Phys. Commun. {\bf 109}, 193 (1998).
\bibitem{Xu11} J. Xu and C. M. Ko, Phys. Rev. C \textbf{84}, 014903 (2011).
\bibitem{art} B. A. Li and C. M. Ko, Phys. Rev. C {\bf 52}, 2037 (1995).

\bibitem{Hos85} A. Hosoya and K. Kajantie, Nucl. Phys. B \textbf{250}, 666 (1985).
\bibitem{Dan85} P. Danielewicz and M. Gyulassy, Phys. Rev. D \textbf{31}, 53 (1985).
\bibitem{Hei96} H. Heiselberg and X. N. Wang, Phys. Rev. C \textbf{53}, 1892 (1996).
\bibitem{Mur02} A. Muronga, Phys. Rev. Lett. \textbf{88}, 062302 (2002); \textbf{89}, 159901(E) (2002); Phys. Rev. C \textbf{69}, 034903 (2004).
\bibitem{De09} N. Demir and S. A. Bass, Phys. Rev. Lett. {\bf
102}, 172302 (2009).

\bibitem{Sie79} P. J. Siemens and J. I. Kapusta, Phys. Rev. Lett. \textbf{43}, 1486 (1979).
\bibitem{Aic87} J. Aichelin and E. A. Remler, Phys. Rev. C \textbf{35}, 1291 (1987).

\bibitem{Dos88} K. G. R. Doss {\it et al.}, Phys. Rev. C \textbf{37}, 163 (1988).
\bibitem{Kuh93} C. Kuhn {\it et al.}, Phys. Rev. C \textbf{48}, 1232 (1993).

\bibitem{Gud85} K. K. Gudima {\it et al.}, Phys. Rev. C \textbf{32}, 1605 (1985).
\bibitem{Ber81} G. Bertsch and J. Gugnon, Phys. Rev. C \textbf{24}, 2514 (1981).

\bibitem{sta} P. Braun-Munzinger, K. Redlich, and J. Stachel, in: R. Hwa and X. N. Wang (Eds.), Quark Gluon Plasma 3, World Scientific, Singapore, 2004, p. 491, arXiv: nucl-th/0304013.
\bibitem{And06} A. Andronic, P. Braun-Munzinger, and J. Stachel, Nucl. Phys. A \textbf{772}, 167 (2006).

\bibitem{Cel06} J. Cleymans, H. Oeschler, K. Redlich, and S. Wheaton, Phys. Rev. C \textbf{73}, 034905 (2006).
\bibitem{And10} A. Andronic, P. Braun-Munzinger, and J. Stachel, Nucl. Phys. A \textbf{834}, 237c (2010).

\end{thebibliography}
\end{document}